\documentclass[11pt,epsf]{article}
\usepackage{epsfig}
\title{{\bf Populations of Dwarfs in Clusters of Galaxies:\\
Environmental Connections\\}}
\author{J. S. Gallagher$^1$, C. J. Conselice$^{1,2}$, R. F. G. Wyse$^3$\\
\vspace{0.1cm}\\
\normalsize $^1$Department of Astronomy, University of
Wisconsin-Madison. Madison, WI, USA\\
\normalsize $^2$Space Telescope Science Institute, 3700 San Martin Drive,
Baltimore MD, USA\\
\normalsize $^3$Department of Physics and Astronomy, Johns Hopkins University,
Baltimore MD, USA\\
}
\date{}
\begin{document}
\maketitle
\def\bull{\vrule height .9ex width .8ex depth -.1ex}
\makeatletter
\def\ps@plain{\let\@mkboth\gobbletwo
\def\@oddhead{}\def\@oddfoot{\hfil\tiny
``Dwarf Galaxies and their Environment'';
Bad Honnef, Germany, 23-27 January 2001; Eds.{} K.S. de Boer, R.-J.Dettmar, U. Klein; Shaker Verlag}%
\def\@evenhead{}\let\@evenfoot\@oddfoot}
\makeatother

%%  if your contribution is short, you may, if the title is clear enough,
%%  skip the abstract.....
\begin{abstract}\noindent
Despite their apparent fragile appearance, dwarf spheroidals
are the most common galaxy type in clusters. In this paper we consider some of
the issues associated with two major models for the origin of these
dwarfs: primeval galaxies which formed with the cluster and the modification
of accreted systems. We argue that the
present observational
evidence, derived from the Virgo and Perseus clusters, points to infall
as the origin of many of these objects.

\end{abstract}

\section{Introduction}

After the advent of high sensitivity photographic emulsions used in
combination with wide field telescopes, it became clear that clusters
of galaxies contain large numbers of low surface brightness dwarf
galaxies (e.g., Binggeli et al. 1985, the Virgo Cluster Catalog (VCC)).  These
consist predominantly of dwarf elliptical and the structurally similar,
but less luminous, dwarf spheroidal systems (hereafter simply `cluster dEs';
e.g., Caldwell 1987, Impey et al. 1988, Driver et al. 1994, Secker et al.
1997). Although dEs appear to be scaled down ellipticals, they fundamentally
differ in some properties such as their mean surface brightnesses within
physical radii which scale with luminosity
in contrast to the inverse correlation in giant elliptical galaxies
(e.g., Wirth \& Gallagher 1984, Kormendy 1985, Ferguson \& Binggeli 1994).

A revolution in cluster dwarf investigations is now in progress driven by
gains from the application of high performance CCD detectors used with
cameras and spectrographs. While the fundamental full-coverage
photographic studies of the Virgo and Fornax clusters by for example
Bingelli et al. (1985) has yet to be equaled, it is clear that
clusters of galaxies routinely display rising luminosity functions to
$M_V \leq -$14, indicative of huge populations of spheroidal dwarfs.
Yet, the origin and evolution of this most common galaxy type
 remains an intriguing problem. Why are the lowest luminosity
density galaxies so numerous in the densest regions of the local
universe? Is this a result of galaxy formation processes in regions
with above average densities, or could they arise from changes in galaxy
evolution induced by interactions within galaxy clusters? We have
undertaken a research program, mainly through optical observations with
the WIYN 3.5-m telescope to explore these and related issues.  The basic
results are contained in Conselice (2001) and are
reported in more detail in Conselice et al. (2001a,b).

 \section{Optical Structures of Cluster Dwarf Galaxies}

Observations with WFPC2 on the {\it
Hubble Space Telescope}, have recently resolved dEs in the Fornax and Virgo
clusters, allowing us to chart their internal properties, especially of
their nuclei and globular star cluster systems (e.g., Miller et al.
1998, O'Neil et al. 1999; Lotz et al. 2001).  Figure~1 shows typical dE
galaxies in the Virgo cluster. These and similar
ground-based observations of nearby galaxy clusters
lead to some fundamental descriptions of cluster dEs:

{\bf (i)} Cluster dEs come in nucleated and
non- nucleated varieties, with the frequency of nuclei increasing with
luminosity (Sandage \& Binggeli 1984, Binggeli \& Cameron 1991,
Ferguson \& Binggeli 1994).
WFPC2 observations confirm suspicions
of ground-based observers that these nuclei are often offset from the
centers of the outer stellar isophotes (cf. Binggeli et al. 2000).
Therefore, unlike most spirals, the
globular cluster-like nuclei of dEs do not always define the centers
of their host systems.

{\bf (ii)} Cluster dEs have a variety of shapes.  This issue has
been most thoroughly explored by Ryden et al. (1999) in the Virgo cluster.
While a few dwarf S0s were identified by Binggeli et al. (1985), Ryden
et al. (1999) found evidence for dEs with both disky
(e.g., VCC 854 in Fig.  1; see also Jerjen et al. 2000) and boxy
isophotal shapes.  The
optical structures of cluster dEs are therefore
inhomogeneous. At an extreme, some
dEs are highly distorted; e.g., near the center of the Coma cluster,
presumably due to tidal disruption (Thompson \& Gregory 1993).

{\bf (iii)} While a basic correlation exists between the optical
colors and luminosities of cluster dEs, the amount of scatter is large,
particularly at low luminosities.  Evidently variations exist in the
age- metallicity relationships between dEs (Cellone \& Forte 1996,
Caldwell \& Rose 1998, Rakos et al. 2001, Conselice et al.  2001b).

Thus, these observations suggest that cluster dEs do not
form a simple population. It is therefore reasonable to consider the
possibility that cluster dEs could originate from more than one
evolutionary path and in complex ways.

\begin{figure}
\begin{center}
\begin{minipage}[][8cm][t]{14cm}
\epsfig{file=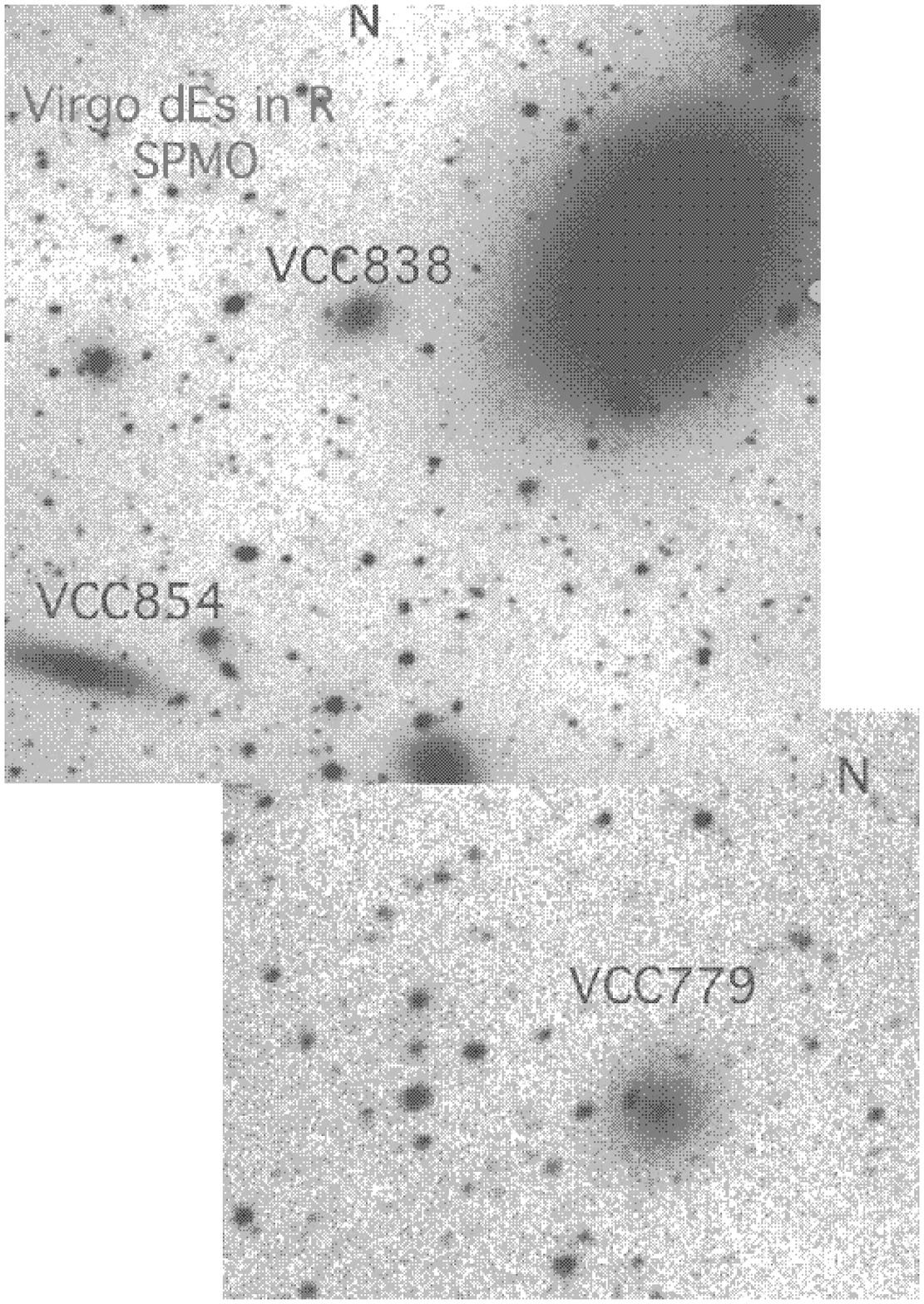,width=6cm}\\
\vskip -2.3in
\hskip 6cm
Figure 1 - Dwarf galaxies in the Virgo

\hskip 6cm cluster labeled with their VCC numbers.

\hskip 6cm From deep R-band images obtained by

\hskip 6cm S. Levine with the 2-m telescope at

\hskip 6cm the San Pedro Martir Observatory.
\vskip 2in
\end{minipage}
\end{center}
\end{figure}
%\vskip 10in
\section{Evolution of Cluster Dwarf Populations}

In this section we develop a basic description of how the total number
of spheroidal dwarfs in a cluster, $N_{dE}(t)$, varies in a simple one
spatial zone toy model.  In so doing we ignore mass-dependent effects,
and any possible non-linearities,
as our goal is only to look at mean properties of moderate luminosity
dEs with $M_V \leq -$14 mag.  If $P_{dE}(t)$ is the production rate of new
dE systems at time $t$ and $D_{dE}(t)$ is the destruction rate, then $
\dot N_{dE}(t) =  P_{dE}(t) - D_{dE}(t)$.  If the cluster formed at
$t_f$, then the number of dEs at the current time $t_0$ is
$$ N_{dE}(t_0) = N_{dE}(t_f) + \int_{t_f}^{t_0}{\dot N_{dE}(t)
dt}. $$
For a purely primordial cluster dE population we
require $P_{dE}(t) = D_{dE}(t) =0$, which is physically unlikely.

More generally, we could expect that dEs are produced from galaxies
infalling into clusters, as in for example the models of
Moore et al.  (1998;
see also Quilis et al. 2000), in which case for an infall rate $\dot
N_c(t)$ we can write $ P_{dE}(t) = c_{dE}(t) \dot N_c(t)$ where
$c_{dE}(t)$ is the efficiency of conversion of infalling galaxies into
dEs, absorbing
any time delay into the time variation of the efficiency.
The destruction rate should depend on a destruction efficiency
$d_{dE}(t)$ such that $ D_{dE}(t) = d_{dE}(t) N_{dE}(t)$.

One model is to assume all dEs are made from galaxies that fall into a
pre-existing cluster.  In this case $ N_{dE}(t_f)=0$, and we can
estimate a minimum infall rate by setting $d_{dE}(t)=0$ for all $t$.
The average infall rate required to make the observed population of dEs
in the Virgo cluster, which we adopt as out standard, is then
$$ \overline{ \dot N_c(t)} = N_{dE}(t_0)/\bar c_{dE} (t_0 - t_f). $$
To get an a rough estimate of the infall rate, we assume a cluster age
of 10~Gyr and a dE production efficiency from all but giant galaxies of
100\%. Then since there are $\sim$10$^3$ dEs in Virgo, we require an average
infall rate of $\geq$100 galaxies per Gyr.  High infall rates in
the past are therefore necessary, but possible (Kauffmann 1995).

Infall rates are difficult to measure, and likely will be episodic
since galaxies may arrive in groups. The number of moderate luminosity
galaxies in Virgo that seem to be in an evolutionary transition is not
clear, but based on Gallagher \& Hunter (1989) it is probably no more than
two dozen currently.  The lifetime of the transition phases is also uncertain,
but from what we now know, it appears that the current infall rate is
too low to produce the observed population of dEs and related objects
in the Virgo cluster, especially if any destruction occurs and the
production efficiency is less than 100\%, as is likely.

An alternative perspective is to assume that all dEs were formed when
the cluster was young. In this case we need to understand the survival rates
of dEs in clusters. Most interactions in clusters occur at
high relative velocities, limiting the damage from any one interaction.
The impulse approximation then holds and the change in
internal energy per collision is given by eq. (7-53) in Binney \& Tremaine
(1987) with the impact parameter set to the size of a giant elliptical,
and hence giving an upper limit on the energy input:

$$ \Delta E/E < 7 (v_{dE}/v_{rel})^2(\bar \rho_{dE}/\bar \rho_{giant}). $$
For typical dEs $ (v_{dE} \leq$ 50~km~s$^{-1}$ while relative
velocities will be $\sim$1000~km~s$^{-1}$) the fractional change in
internal energy per collision is $ \Delta E/E <$10$^{-3}$ and with only
a few collisions per cluster crossing, most dEs will survive collisional
heating in a more or less intact state, while more massive galaxies
with larger internal velocities will tend to be worn down due to their
stronger reactions to collisions. Similarly, the description of tidal
disruption by Merritt (1984) shows that most of the stellar bodies of dEs can
survive, with only the least massive systems being subject to disruption
near cluster cores (cf. Thompson \& Gregory 1993, Adami et al. 1998),
or around central cluster giant galaxies (L\'opez-Cruz et al. 1997).

We conclude that in Virgo, $ D_{dE}(t_0)$ is small, and that the infall
rate of smaller galaxies exceeds the destruction rate of dEs.  The
production rate of dEs is less clear due to uncertainties in the
astrophysics of converting field galaxies to cluster dEs, but we
suspect $P_{dE}(t_0) > D_{dE}(t_0)$ and that therefore
the dE population of Virgo is increasing.

\section{A Kinematic Test of the Infall Model}

While the cluster infall model for making dE galaxies has received
theoretical and observational support, sharp tests are difficult. Most
of the accretion is expected to have occurred several Gyr in the past
(Kauffmann 1995), and it is difficult to observe accretion in clusters
at that look back time which corresponds to redshifts $z \sim 1$
(but see Martin et al.
2000).  One signature of infall would be a range of time scales for the
cessation of star formation in dEs, but this is also found in Local
Group dEs and dSph, and is challenging to measure in cluster dwarfs due
to the effects of the well-known age-metallicity degeneracy on the
integrated light of middle age or older stellar populations. We have
adopted a third approach for measuring ages: the examination of global
kinematics of the Virgo cluster dE population.

\begin{figure}
\begin{center}
\begin{minipage}[][9.5cm][t]{8cm}
\epsfig{file=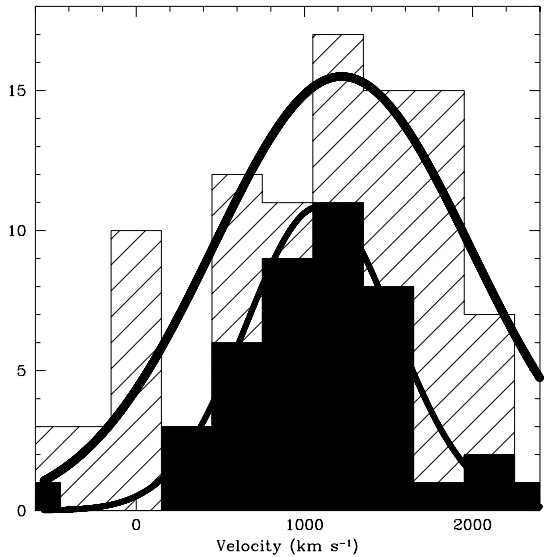,width=8cm}\\
\vskip -0.85cm
\caption{Velocity distribution for the elliptical (solid histograph) and
dwarf ellipticals (hatched histograph).  Solid lines show the formal
Gaussian fits to each distribution.}
\end{minipage}
\end{center}
\end{figure}

 A virialized object such as a galaxy cluster grows in density contrast
over time with respect to a surrounding expanding universe. Depending
on $\Omega_{\rm matter}$ and $\Omega_{\lambda}$,  the cluster
will accrete matter from its surroundings at various rates, with
cosmologically expanding matter shells turning around under the
influence of gravity (Gunn \& Gott 1972; Hamilton 2001). As this
process proceeds, the cluster's original core gets buried in a larger
system.  In principle we then can use the kinematic characteristics of
cluster members to define an accretion sequence. Old fully virialized
objects will have a peaked, Gaussian-like radial velocity distribution,
while objects that now are infalling will have a flat distribution of
velocities (Huchra 1985, Schindler et al. 1999).  Objects recently
acquired by a cluster should lie between these two extremes.

To answer this question we undertook a program to measure radial velocities
of Virgo cluster
dEs galaxies with the Hydra multi-object fiber system feeding the bench
spectrograph on the WIYN Telescope
(see Conselice et al. 2001a for
details).  Combining our new velocities with a larger set from the
literature (e.g., Bothun \& Mould 1988, Schindler et al. 1999), we
found that the dE members of Virgo have intermediate kinematic
properties between those of the `old core' giant elliptical galaxies, and
infalling irregular and spiral systems (Figure 2). The velocity distribution of
the dEs is significantly wider than that of the E galaxies ($\sigma
\approx$ 730~km~s$^{-1}$ vs. 460~km~s$^{-1}$). Furthermore, the dEs
display more spatial and velocity substructure (Figure 2).
The combination of large and substructured
velocity and spatial distribution of these dEs, and
the fact that the velocity distribution of Virgo dEs is not well fit by a
Gaussian and has a ratio with the elliptical galaxies expected for a
virialized and
accreted component all suggest that dEs are relatively recent additions to
the cluster (Conselice et al. 2001a).  This also implies that the dEs have an
intermediate cluster dynamical age.  Thus, many Virgo dEs are not left over
from the initial cluster formation.

\section{Discussion}

Our results lead us to a model where Virgo (and other galaxy cluster) dEs arose
from a variety of processes, their spheroidal shapes resulting at least
in part from combined effects of gas stripping (Mori \& Burkert 2000)
and dynamical heating (Moore et al. 1998).  The integrated colors of
dEs indicate that major star formation typically ceased at least
$\sim$3-5~Gyr in the past. Combining this with the Virgo kinematic
results shows that most dEs are likely to be old, but not ancient
members of clusters.  Thus, a cluster membership age spread
probably exists among Virgo dEs. Ferguson and Sandage (1989) noted that
the nucleated dEs in Virgo are more centrally concentrated around the
giant Es than are the diffuse dEs, which could reflect a difference in
age.  Unfortunately the kinematic data are not yet sufficient to
rigorously check for kinematic differences between the various
subclasses of Virgo cluster dEs.

We do not yet know the forms of cluster dEs at birth.  The harassment
mechanism suggests that moderate mass galaxies, that in the field are
small spirals or large irregular galaxies, are good prospects for
becoming dEs once they are captured by a galaxy cluster.  Sandage and
Binggeli (1984) noted that these types of galaxies are common in the
field but deficient in galaxy clusters (see also Binggeli et al. 1985);
therefore it is possible that cluster versions of these
objects have morphologically evolved.  Our empirical
arguments, and other theoretical models suggest that the galaxy infall rate in
clusters peaked about 5-10~Gyr in the past, and this could be when many
present-day dEs appeared in the cluster.  At these times the conversion of a
star-forming disk galaxy into a dE may have been easier since less of a
late-type galaxy's baryonic mass would have been converted into stars
and thus the effects of gas stripping would have been more dramatic.

If the views presented here prove correct, then dwarf spheroidal
galaxies in clusters are substantial products of galaxy evolution rather than
primordial `formation'.  This argues against the identification of
dE/dSph systems as a
class with a first generation of violently star-forming dwarf galaxies,
as are expected in some cold dark matter galaxy formation models.  We
can look forward to eventually resolving this through determinations of
the stellar population age and metallicity spreads in cluster and field
dEs, an effort that now is under way for nearby systems and can, with
30-m class telescopes operating near their diffraction limit in the
near infrared, be extended to the Virgo and Fornax clusters.

\vspace*{0.2cm}

We wish to thank the National Science Foundation for support of this
research through grants AST9803018 to the University of Wisconsin-Madison and
AST9804706 to Johns Hopkins University. JSG also thanks the Vilas Trust
for their generous grant through the University of Wisconsin-Madison
Graduate School. CJC thanks NASA for a Graduate Student Research
Program Fellowship.
We also express our appreciation
to the WIYN Observatory crew for making our quality observations
possible and S. Levine for Figure 1.

% References:
%%  Use the ApJ, AJ, new A\&A style (are the same!)
\newpage
{\small
\begin{description}{} \itemsep=0pt \parsep=0pt \parskip=0pt \labelsep=0pt
\item {\bf References}

\item
Adami, C. et al. 1998, A\&A, 334, 765
\item
Binggeli, B. \& Cameron, L. M. 1991, A\&A, 252, 27
\item
Binggeli, B., Binggeli, B., \& Tammann, G.A. 1985, AJ, 90, 395
\item
Binggeli, B., Barazza, F., \& Jerjen, H. 2000, A\&A, 359, 447
\item
Binney, J., \& Tremaine, S. 1987, ``Galactic Dynamics'', Princeton University Press, Princeton, N.J.
\item
Bothun, G.D., \& Mould, J.R. 1988, ApJ, 324, 123
\item
Caldwell, N. 1987, AJ, 94, 1116
\item
Caldwell, N. \& Bothun, G. D. 1987, AJ, 94, 1126
\item
Caldwell, N. \& Rose, J. A. 1998, AJ, 115, 1423
\item
Cellone, S. A. \& Forte, J. C. 1996, ApJ, 461, 176
\item
Conselice, C.J. 2001, PhD Thesis, University of Wisconsin-Madison
\item
Conselice, C.J., Gallagher, J.S., \& Wyse, R.F.G. 2001a, ApJ, in press, astro-ph/0105492
\item
Conselice, C.J., Gallagher, J.S., \& Wyse, R.F.G. 2001b, in prep.
\item
Driver, S. P., Phillipps, S., Davies, J. L., Morgen, I., \& Disney,
M. J. 1994, MNRAS, 268, 393
\item
Ferguson, H. C. \& Sandage, A. 1989, ApJ, 346, L53
\item
Ferguson, H. C. \& Binggeli, B. 1994, A\&ARv, 6, 67
\item
Gallagher, J. S. \& Hunter, D. A. 1989, AJ, 98, 806
\item
Gunn, J. E. \& Gott, III, J. R. 1972, ApJ, 176, 1
\item
Hamilton, A.J.S. 2001, MNRAS, 322, 419
\item
Huchra, J. P. 1985 in The Virgo Cluster, ESO Workshop Proceedings 20,
eds. O. G. Richter \& B. Binggeli, p.181
\item
Impey, C., Bothun, G. \& Malin, D. 1988, ApJ, 330, 634
\item
Jerjen, H., Kalnajs, A., \& Binggeli, B. 2000, A\&A, 358, 845
\item
Kauffmann, G. 1995, MNRAS, 274, 143
\item
Kormendy, J. 1985, 295, 73
\item
L\'opez-Cruz, O., Yee, H. K. C., Brown, J. P., Jones, C., \& Forman, W. 1997,
ApJ, 475, L97
\item
Lotz, J.M., Telford, R., Ferguson, H.C., Miller, B.W., Stiavelli, M., \& Mack, J. 2001, ApJ, 552, 572
\item
Martin, C.L., Lotz, J., \& Ferguson, H. C. 2000, ApJ, 543, 97
\item
Merritt, D. 1984, ApJ, 276, 26
\item
Miller, B. W., Lotz, J. M., Ferguson, H. C., Stiavelli, M.,
\& Whitmore, B. C. 1998, ApJ, 508, L133
\item
Moore, B., Lake, G. \& Katz, N. 1998, ApJ, 495, 139
\item
Mori, M. \& Burkert, A. 2000, ApJ, 538, 559
\item
O'Neil, K., Bothun, G. D., \& Impey, C. D. 1999, AJ, 118, 1618 (Erratum
1999, AJ, 119, 984)
\item
Quilis, V., Moore, B., \& Bower, R. 2000, Science, 288, 1617
\item
Rakos, K., Schombert, J., Maitzen, H. M., Prugovecki, S., \&
Odell, A. 2001, AJ, 121. 1974
\item
Ryden, B.S., Terndrup, D.M., Pogge, R.W., \& Lauer, T.R. 1999, ApJ, 517, 650
\item
Sandage, A. \& Binggeli, B. 1984, AJ, 89, 919
\item
Secker, J., Harris, W.E., \& Plummer, J.D. 1997, PASP, 109, 1377
\item
Schindler, S., Binggeli, B., \& Bohringer, H. 1999, A\&A, 343, 420
\item
Thompson, L. A. \& Gregory, S. A. 1993, AJ, 106, 2197
\item
Wirth, A. \& Gallagher, J. S. 1984, ApJ, 282, 85

\end{description}
}

\end{document}